\documentclass{aa}  
\usepackage{graphicx}
\usepackage{amsmath}
\usepackage{epstopdf}
\usepackage[]{natbib}
\usepackage{multirow}
\usepackage{booktabs}
\usepackage{txfonts}
\usepackage{bigstrut}
\usepackage{enumitem}   
\begin{document}
   \title{Gamma rays from clumpy wind-jet interactions in high-mass microquasars} 

   \titlerunning{Clumpy winds in microquasars}
   \authorrunning{V. M. de~la~Cita, S. del~Palacio, V. Bosch-Ramon, et al.} 

   \author{
   V. M. de~la~Cita \inst{1}
   \and S. del~Palacio \inst{2,3,} \thanks{Fellow of CONICET}
   \and V. Bosch-Ramon \inst{1}
   \and X. Paredes-Fortuny \inst{1} 
   \and G. E. Romero \inst{2,3}
    \and D. Khangulyan\inst{4}
    }

   \institute{
   Departament d'Astronomia i Meteorologia, Institut de Ci\`ences del Cosmos (ICCUB), Universitat de Barcelona (IEEC-UB), Mart\'{\i} i Franqu\`es 1, E-08028 Barcelona, Spain
    \and Instituto Argentino de Radioastronom\'{\i}a (CCT La Plata, CONICET), C.C.5, (1894) Villa Elisa, Buenos Aires, Argentina.
    \and Facultad de Ciencias Astron\'omicas y Geof\'{\i}sicas, Universidad Nacional de La Plata, Paseo del Bosque, B1900FWA 
La Plata, Argentina.
    \and Department of Physics, Rikkyo University 3-34-1, Nishi-Ikebukuro, Toshima-ku, Tokyo 171-8501, Japan
    }

   \date{Received - ; accepted - }

  \abstract
   {The stellar winds of the massive stars in high-mass microquasars are thought to be inhomogeneous. The interaction of these inhomogeneities, or clumps, with the jets of these objects may be a major factor in gamma-ray production.}
   {Our goal is to characterize a typical scenario of clump-jet interaction, and calculate the contribution of these interactions to the gamma-ray emission 
   from these systems.}
   {We use axisymmetric, relativistic hydrodynamical simulations to model the emitting flow in a typical clump-jet interaction. Using the simulation results we perform a numerical calculation of the high-energy emission from one of these interactions. The radiative calculations are done for relativistic electrons locally accelerated at the jet shock, and the synchrotron and inverse Compton radiation spectra are computed for different stages of the shocked clump evolution. We also explore different parameter values, such as viewing angle and magnetic field strength. We generalize phenomenologically the results derived from one clump-jet interaction to multiple interactions under different wind models, estimating the clump-jet interaction rates, and the resulting luminosities in the GeV range.}
   {If particles are efficiently accelerated in clump-jet interactions, the apparent gamma-ray luminosity through inverse Compton scattering with the stellar photons can be significant even for rather strong magnetic fields and thus efficient synchrotron cooling. Moreover, despite the standing nature or slow motion of the jet shocks for most of the interaction stage, Doppler boosting in the postshock flow is relevant even for mildly relativistic jets.}  
   {For clump-to-average wind density contrasts $\gtrsim 10$, clump-jet interactions could be bright enough to match the observed GeV luminosity in Cyg~X-1 and Cyg~X-3 when a jet is present in these sources, with required non-thermal-to-total available power fractions $\gtrsim 0.01$ and $0.1$, respectively.}
   \maketitle
%


\section{Introduction}\label{sec:intro}


High-mass microquasars (HMMQs) are binary systems hosting a massive star, and a compact object able to produce jets in which high-energy
(HE) processes can take place \citep[see, e.g.,][and references therein]{BosKha09,Dub13,Bed13}. To date, gamma rays have been robustly detected from two HMMQs, Cyg~X-3 and Cyg~X-1 \citep[][respectively]{tbp09,zf16}. Variability of the detected emission indicates that the high-energy source should be located relatively close to the compact object, at a distance comparable to the binary separation distance \citep{dch10*a,zf16}. There is a tentative detection of another HMMQ, SS~433, although in this case the emission would be likely coming from the jet-termination region \citep{byk15}.

The energies at which Cyg~X-1 and Cyg~X-3 were detected are in the GeV range, except for a flare-like detection of Cyg~X-1 with the MAGIC Cherenkov telescope in the TeV range with post-trial significance of $4.1\sigma$ \citep{aaa07}. Both sources present long-term gamma-ray emission, overlapping with possible day-scale flares, all associated with jet
activity \citep[][]{aaa07,tbp09,faa09,sts10,mzc13,btp13,zf16,zsp16,zmc16}. In what follows, we will carry a mixed approach considering HMMQs in general, while adopting Cyg~X-1 and Cyg~X-3 as reference sources to check our results in the context of real objects.

Massive stars produce dense and fast winds that are thought to be inhomogeneous \citep[e.g.][and references therein]{RunOwo02,Mof08,pvn08}. The
specific properties of these inhomogeneous winds may depend on the stellar type and evolutionary phase, but in general they can be 
described in terms of dense clumps in a dilute medium. In particular, Cyg~X-1 hosts a black hole and an O-type supergiant, and Cyg~X-3 either a neutron star or a black hole, and a Wolf-Rayet star, and clumpy winds have been suggested to be present in both systems \citep[see, e.g.][]{SzoZdz08,rlh11,mhh16}. 

Detection of gamma-ray emission from galactic jet sources requires efficient particle acceleration, which is conventionally associated with shocks. The propagation of a jet through the binary system environment may come along with the formation of shocks at binary scales in addition to the jet termination shock. For example, internal shocks form when portions of jet material, moving with different velocities, collide to each other \citep[e.g.][]{brp06}. Furthermore, the stellar wind lateral impact should produce asymmetric recollimation shocks and induce non-thermal emission
\citep[e.g.][]{rtk03,PerBos08,dch10,yzh16}. But it cannot be neglected that when wind density inhomogeneities or clumps penetrate inside the jets, they should trigger strong shocks as well. Thus, the wind clumps in HMMQs may have a significant influence on the jet dynamics
\citep{PerBos12} and on the non-thermal HE processes occurring on the scales of the binary system
\citep{ort09,abr09,rdo10}. Therefore, despite all these shocks can be sites of efficient particle acceleration and HE emission, one expects the strongest kinetic-to-internal energy conversion for shocks associated with wind clumps present inside the jet \citep{Bos15}. Moreover, besides this high-conversion efficiency, the Doppler boosting of the non-thermal emission associated with wind clumps, in addition to being important, might be also favourable for relatively off-axis observers. 

In this work, we present for the first time numerical calculations of the HE emission produced by a clump-jet interaction in a HMMQ using the hydrodynamical information obtained from a simplified relativistic, hydrodynamical (RHD) axi-symmetric simulation of such an interaction. Particle acceleration is assumed to occur in the jet shock, as this is much more energetic than the one initially crossing the clump. The parameters adopted have been chosen such that the simulation can be taken as a reference case, i.e. a clump of realistic parameters being inside the jet at a typical interaction jet height, similar to the binary size. Interactions taking place significantly closer, or farther, from the jet base, would be either very unlikely, or too oblique for the clump to penetrate into the jet  (see Eq.~\ref{eq:clump_min}). Therefore, the calculations performed can be used as a reference to establish the typical radiation outcome from one interaction, choosing also suitably the orbital phase, although some of the results have been checked for different orbital phases. The radiation results can be then generalized to the realistic case of multiple wind clumps interacting with the jet, adopting a phenomenological prescription for the inhomogeneous wind properties, similarly to what was done in \cite{Bos13} in the context of a high-mass binary hosting a young pulsar.

The paper is organized as follows: In Sect.~\ref{sec:scenario}, the conditions under which a clump is able to penetrate into a jet are computed. Then, in Sect.~\ref{sec:simulation}, first, we briefly introduce the results of a simplified RHD axi-symmetric simulation of a characteristic clump-jet interaction, and the streamline approach to define the emitter structure; next, we introduce the non-thermal radiation calculations and their results. In Sect.~\ref{sec:results}, a generalization of the numerical radiation results is carried out adopting a phenomenological prescription for the clumpy wind properties. Finally, a discussion of the obtained results in the context of the HMMQs Cyg~X-1 and Cyg~X-3 is presented in Sect.~\ref{sec:disc}, along with a summary of the conclusions of the work.

\section{Clump-jet interaction: basic estimates}\label{sec:scenario}

In this section we study analytically the main characteristics of a clump-jet interaction, and show that even for 
conservative assumptions it is likely that clumps penetrate the jet. The following quantities are required for the analysis: (i) the stellar mass-loss rate ($\dot{M}_\mathrm{w}$), wind velocity ($v_\mathrm{w}$), and wind density ($\rho_\mathrm{w}$); (ii) the jet luminosity (without accounting for the rest-mass energy; $L_\mathrm{j}$), velocity ($v_\mathrm{j}$, or Lorentz factor $\Gamma_\mathrm{j}$), radius ($R_\mathrm{j}$), height ($z_\mathrm{j}$), density ($\rho_\mathrm{j}$), and jet half opening angle ($\theta_\mathrm{j}=R_\mathrm{j}/z_\mathrm{j}$); (iii) the clump characteristic radius ($R_\mathrm{c}$), and density ($\rho_\mathrm{c}$) (assuming spherical uniform clumps), which relates to the average wind density through the density contrast or clumping factor ($\chi=\rho_\mathrm{c}/\rho_\mathrm{w} > 1$); and finally (iv) the distance between the star and the base of the jet, or orbital separation distance ($R_\mathrm{orb}$). In this work, the jet is assumed to be perpendicular to the orbital plane. We also consider a mildly relativistic jet (see Sect.~2 in \citealt{BosBar16}), with $\Gamma_\mathrm{j}=2$. The hydrodynamic approximation for the clump-jet is adopted, as the gyroradii of the particles involved are much smaller than the typical size of the interacting structures.

We consider the emission of a clump interacting with the jet at a distance similar to $R_\mathrm{orb}$. On these scales, clump-jet interactions are more numerous because the jet is thicker; the jet is also more dilute than further upstream, and the clump velocity is still rather perpendicular to the jet, favouring jet penetration. Moreover, at smaller distances from the compact object, the jet ram pressure is too strong for the clumps to survive penetration.
For high density contrast values, i.e. $\chi\gg 1$, wind-jet interactions may occur just through clumps entering into the jet. In that case, the clumps would be surrounded by a very dilute, hot medium of little dynamical impact either for the jet, or the clumps. We adopt here however the more conservative assumption that, even if $\chi\sim 10$, the wind interacts with the jet forming a relatively smooth region of shocked material that circumvents the jet (see \citealt{PerBos12}; see also \citealt{Pit07} in the context of colliding wind binaries). This shocked wind can prevent small clumps from reaching the jet, which sets the first condition for clump-jet interaction to occur. 
Moreover, the impact of the jet in a clump generates a shock that propagates in the latter, eventually destroying it. Therefore, a 
second condition is that the forward shock in the clump is slower than the clump velocity perpendicular to the jet, to allow the clump to enter deep enough into the jet before its destruction \citep[see, e.g.][]{abr09}.

To formalize the first condition, let us consider a clump that travels with velocity $\sim v_\mathrm{w}$ perpendicular to the jet and reaches the region where the wind interacts with the jet boundary. To successfully penetrate the jet, the clump has to go through the shocked wind, with respect to which the clump is moving also at $\sim v_{\rm w}$, without significantly slowing down. Such a region has a thickness $\sim R_\mathrm{j}$ and exerts a drag on the clump that can be quantified through a ram pressure $P_\mathrm{w}\sim \rho_\mathrm{w} v_\mathrm{w}^2$.
The acceleration exerted on the clump, which has a characteristic surface 
$s_\mathrm{c}\sim \pi R_\mathrm{c}^2$  
and volume 
$V_\mathrm{c}\sim 4/3\,s_\mathrm{c}\,R_\mathrm{c}\sim s_\mathrm{c}\,R_\mathrm{c}$, 
by this pressure is 
$a_\mathrm{acc}\sim s_\mathrm{c}\, P_\mathrm{w}/m_\mathrm{c}$. 
Using that 
$m_\mathrm{c} = V_\mathrm{c} \, \rho_\mathrm{c}$, 
$V_\mathrm{c}/s_\mathrm{c}\sim R_\mathrm{c}$, 
and 
$\rho_\mathrm{c} = \chi \rho_\mathrm{w}$, 
one can obtain the expression: 
$a_\mathrm{acc} \sim P_\mathrm{w}/(\chi \rho_\mathrm{w} R_\mathrm{c})$. Therefore, the typical 
distance required to significantly slow down the clump in the shocked wind surrounding the jet is 
$l\sim v_\mathrm{w}^2/a_\mathrm{acc} \sim \chi R_\mathrm{c}$, 
which as noted above should be 
$l\gtrsim R_\mathrm{j}$, 
and therefore 
$R_\mathrm{c}\gtrsim R_\mathrm{j}/\chi$,
setting a lower limit on the clump size:
\begin{equation} \label{eq:clump_min}
R_\mathrm{c} > R_0=3\times10^{10} \left(\frac{\theta_\mathrm{j}}{0.1}\right) \left(\frac{10}{\chi}\right) \left(\frac{R_\mathrm{orb}}{3\times10^{12}\,\mathrm{cm}}\right) \,\mathrm{cm}.
\end{equation}

The second condition for clumps to fully penetrate into the jet at $z_{\rm j}\lesssim R_\mathrm{orb}$ is $v_\mathrm{sh}\lesssim v_\mathrm{w}$ 
\footnote{A more precise relation is $v_\mathrm{sh} < v_\mathrm{w \perp}$, but for simplicity in this analysis we assume that $v_\mathrm{w \perp}\sim v_\mathrm{w}$ in the regions of interest.}, 
where $v_\mathrm{sh}$ is the velocity of the shock produced within the clump by the jet impact. 
For a cold jet, its pressure $P_\mathrm{j}$ is dominated by the kinetic component of the 
momentum flux, i.e. $P_\mathrm{j} \sim \Gamma_\mathrm{j}^2 \rho_\mathrm{j} v_\mathrm{j}^2$. Assuming  equilibrium between the jet ram pressure and shocked clump pressure, we obtain $v_\mathrm{sh} \sim (P_\mathrm{j}/\rho_\mathrm{c})^{1/2} \sim \Gamma_\mathrm{j}(\rho_\mathrm{j}/\rho_\mathrm{c})^{1/2} v_\mathrm{j}$. Considering that
 \begin{align}
 \rho_\mathrm{j} &= \frac{L_\mathrm{j}}{\pi R_\mathrm{j}^2 \Gamma_\mathrm{j} (\Gamma_\mathrm{j}-1) v_\mathrm{j}c^2}\,, \\
 \rho_\mathrm{c} &\sim \frac{\chi \dot{M}_\mathrm{w}}{4 \pi R_\mathrm{orb}^2 v_\mathrm{w}}\,,
 \end{align}
and fixing $z_\mathrm{j}\sim R_\mathrm{orb}$, i.~e. $R_\mathrm{j} \sim \theta_\mathrm{j} R_\mathrm{orb}$, one gets the following limitation for the jet power:
\begin{multline} \label{eq:Ljet}
L_\mathrm{j}\lesssim 1.4\times 10^{37} \, \left(\frac{\Gamma_\mathrm{j}-1}{\Gamma_\mathrm{j}}\right) 
\left(\frac{\chi}{10}\right)
\left(\frac{\dot{M}_\mathrm{w}}{3\times 10^{-6} M_\odot \mathrm{yr}^{-1}}\right)\\
\left(\frac{\theta_\mathrm{j}}{0.1}\right)^2
\left(\frac{v_\mathrm{w}}{10^8 \mathrm{cm \, s}^{-1}}\right) 
\left(\frac{v_\mathrm{j}}{c}\right)^{-1} \mathrm{erg} \, \mathrm{s}^{-1}.
\end{multline}
For a hot jet, the specific enthalpy should be included to derive $P_\mathrm{j}$.

For the typical parameters of Cyg~X-3 and Cyg~X-1 \citep[see][and references therein]{yzh16,BosBar16}, and adopting $\chi\sim 10$ and $\Gamma_{\rm j}\sim 2$, clumps will be able to penetrate into the jets of these HMMQ for $R_{\rm c}\sim 3\times 10^9$ and $3\times 10^{10}$~cm, and $L_{\rm j}\lesssim 10^{38}$ and $10^{37}$~erg~s$^{-1}$, respectively. These jet powers are actually similar to the values estimated for these sources, although the uncertainties in the clumpy wind and the jet properties are large.

\section{Clump-jet interaction: numerical calculations}\label{sec:simulation} 

\subsection{Hydrodynamics}

  \begin{figure*}
    \centering
    \includegraphics[width=1.02\textwidth, angle=0]{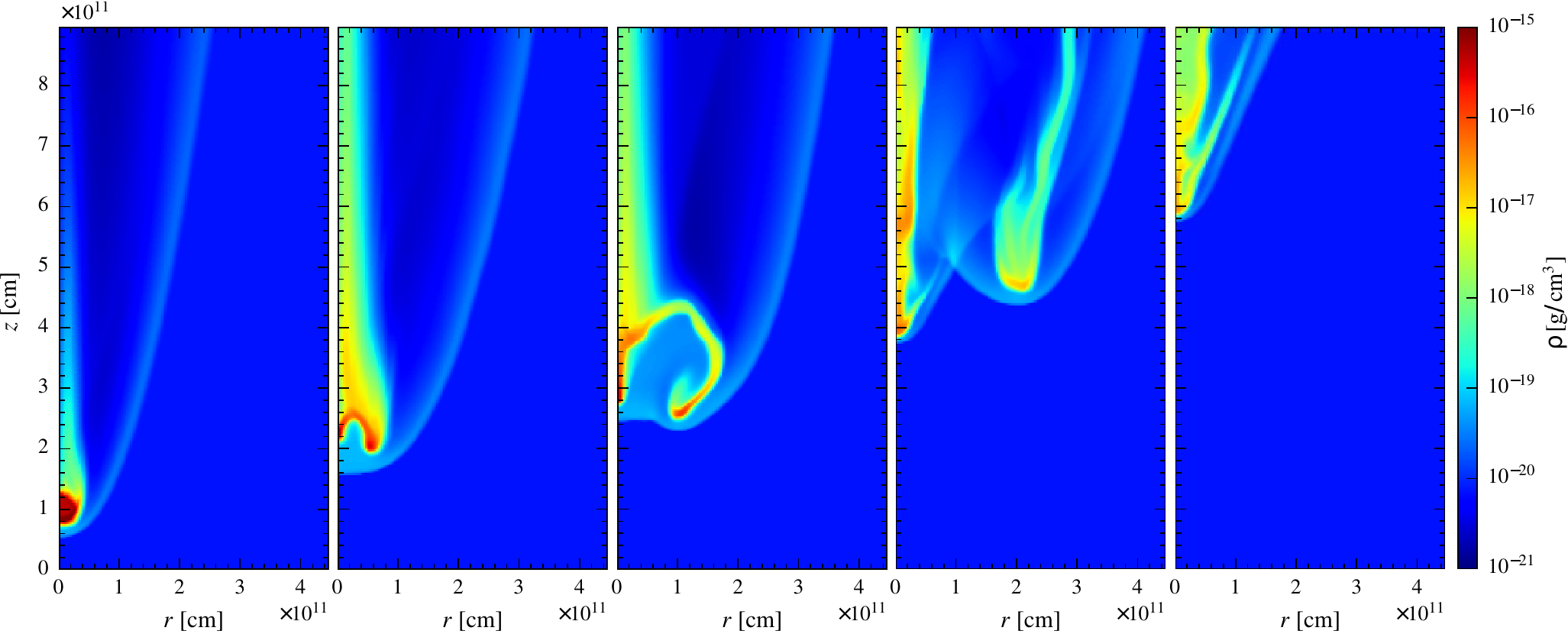}
    \caption[]{Density maps of the clump-jet interaction simulation that illustrate the clump evolution after $\approx 73$, 521, 596, 708, and 822~s (or, equivalently, to $\approx 0.5$, 3.5, 4.1, 4.8, and  5.6 in $R_{\rm c}/v_{\rm sh}$ units).}
    \label{fig:dens}
  \end{figure*}  
  
The clump-jet interaction was simulated in two dimensions (2D) assuming axisymmetry, i.e. neglecting the clump motion with respect to, say, the compact object frame, and a dynamically negligible magnetic field. An adiabatic gas with constant index $\hat\gamma=4/3$ was assumed\footnote{The adiabatic index is constant in the code, although adopting $\hat\gamma=5/3$ yields similar hydrodynamical results. The $\hat\gamma$-value also affects quantitatively the non-thermal processes as they depend on the kinetic-to-internal energy conversion, although we consider the related lack of precision acceptable given the numerous simplified assumptions adopted.}. The RHD equations were solved using the Marquina flux formula \citep{DonMar96,dfi98}. The code is the same as the one used in \cite{Bos15}, but the spatial reconstruction scheme has been improved from 2nd to 3rd order applying the piecewise parabolic method reconstruction scheme \citep{ColWoo84,MarMul96,mpb05}.

The resolution of the calculations is 300 cells in the vertical direction, the $z$-axis, and 150 in the radial direction, the $r$-axis, and the physical size is $z_{\rm grid}^{\rm max}=9\times 10^{11}$~cm in the $z$-direction, and $r_{\rm grid}^{\rm max}=4.5\times 10^{11}$~cm in the $r$-direction. The resolution was chosen such that no significant differences could be seen in the hydrodynamical results when going to higher resolution simulations. Inflow conditions (the jet) are imposed at the bottom of the grid, reflection at the axis, and outflow in the remaining grid boundaries. On the scales of the grid, for simplicity we approximate the jet streamlines at injection as oriented along the $z$-direction despite the jet is actually assumed to be conical.

The injected jet power without accounting for the jet rest-mass energy is $\approx 2.3\times 10^{37}$~erg~s$^{-1}$ for the whole grid, up to $r_{\rm grid}^{\rm max}$, with a Lorentz factor $\Gamma_{\rm j}=2$. The initial clump radius is $R_{\rm c}=3\times 10^{10}$~cm, and its density, $\rho_{\rm c}\approx 5\times 10^{-14}$~g~cm$^{-3}$. This density would correspond to that of a clump located at $z_{\rm j}\approx 3\times 10^{12}$~cm, for a stellar wind with $\dot{M}\approx 3\times 10^{-6}$~M$_\odot$~yr$^{-1}$, $v_{\rm w}\approx 2\times 10^8$~cm~s$^{-1}$, and $\chi\approx 10$ (plus $R_{\rm orb}=3\times 10^{12}$~cm). The initial clump location in the grid is $(0,10^{11}\,{\rm cm})$. 

With the jet impact, the clump gets shocked, expands, disrupts, and eventually leaves the grid \citep[see, e.g.,][in the context of extragalactic jets]{bos12}. The whole duration of the simulation was $\approx 820$~s. Five snapshots of the density distribution, illustrating the clump evolution after $\approx 73$, 521, 596, 708, and 822~s, are presented in Fig.~\ref{fig:dens}. The shocked jet material is seen in light blue, forming a sort of cometary tail pointing upwards and surrounding the clump shocked material, red with green borders.
  
The grid size has been chosen such that the simulation captures the first stages of the clump-jet interaction. This is enough to compute the non-thermal emission for two typical instances of the shocked clump evolution: (i) a quasi-stationary shock in the jet flow is present, but the clump has not expanded, nor it has been displaced, due to the jet impact; (ii) the clump has already expanded, disrupted, and moved along the jet axis. 
A simulation with a significantly larger computational grid, and a much longer simulated time, are required for an accurate description of the clump-jet interaction until the clump has reached an asymptotic speed (and probably fully fragmented and spread in the jet), i.e. when no shock is present in the jet flow (or it is much weaker). This should not affect qualitatively the high-energy emission predictions, but quantitative differences are expected. An accurate study of this (clump-jet mixing, lightcurves, etc.) is left for future work.
The presence of the magnetic field, or 3D calculations, are likely to introduce further complexity to the problem through effects such as suppression, anisotropy, or enhancement, of the growth of instabilities. These effects should be thoroughly studied through devoted simulations, although such an investigation is out of the scope of the present work and left for the future. Other effects may be included also for accuracy, as a more realistic equation of state, or the back-reaction effects of non-thermal processes in the (magneto)hydrodynamics.

\subsection{Streamlines}

To compute the injection, evolution and radiation of the non-thermal particles, the shocked jet flow is modelled as a set of streamlines, each one divided in several cells. These streamlines correspond to the trajectory followed by a fluid element in the flow, which is assumed to be stationary. The assumption of stationarity for the shocked jet flow is valid because the dynamical time of the clump-jet interaction is much longer than the grid crossing time of this flow. The details of the streamline properties are described in \cite{bp16}, Appendix A. The distribution of the streamlines in the radial direction in the present calculations is illustrated in Fig.~\ref{fig:dop}. 

Only streamlines injected at $z_{\rm grid}=0$ and $r_{\rm grid}\le 3\times 10^{11}$~cm are considered for radiation calculations. The reason is that the jet radius at the interaction location is assumed to be $R_{\rm j}=3\times 10^{11}$~cm, which is equivalent to adopting a jet interaction height at $z_{\rm j}\approx R_{\rm orb}$ and a half opening angle $\theta_{\rm j}\approx 0.1$~rad. The actual jet flow injection in the computational grid takes place however up to $r_{\rm grid}^{\rm max}$ in the $r$-axis. This was done for simplicity, as the actual jet surroundings may be very complex because of the stellar wind presence and the wind-jet interaction asymmetry. Thus, introducing a jet boundary in the hydrodynamical simulation seemed to us somewhat artificial if its conditions could not be set realistically. The total injected power in the streamlines (again without taking into account the jet rest-mass energy), i.e. within $R_{\rm j}=3\times 10^{11}$~cm, is $\approx 10^{37}$~erg~s$^{-1}$.

Given the cylindrical symmetry of the simulation, each one of the cells in which the streamlines are divided represents an annular element of fluid in terms of volume and therefore energy content, but from the point of view of particle energy evolution, propagation, and radiation, the cells are to be treated as point-like, as all these processes are not symmetric with respect to the jet axis. In order to mimic the 3D structure of the whole emitter, we assign a random azimuthal angle $\psi$ to each streamline when computing radiation, letting it vary from 0 to $2\pi$ radians to cover all the orientations \citep[see][]{bp16}. 

\subsection{Non-thermal radiation computation}

We need the fluid information for each cell in each streamline to compute where and how much non-thermal particle energy is injected in the flow, and the evolution of the injected particles and later the inverse Compton (IC) and synchrotron radiation. We neglect at this point hadronic processes and relativistic Bremsstrahlung, as synchrotron and IC are in general far more efficient in terms of radiated energy \citep[see, e.g.,][]{BosKha09}. Each streamline is divided in 200 cells, enough to properly sample the flow, with a number of parameters: position and velocity vectors, pressure ($P$), density ($\rho$), section ($S$), magnetic field ($B$), and the fluid velocity divergence ($\,{\rm div} ({\Gamma\mathrm{v}})\,$), needed for the computation of adiabatic losses. For simplicity, the $B$-value in the fluid frame (FF) is computed assuming to be perpendicular to the flow velocity, and that a fraction $\chi_B$ of the total flow energy flux is in the form of Poynting flux \citep{bp16}; two values are adopted: $\chi_B=10^{-3}$ and $1$. Note that the latter value is formally inconsistent with the hydrodynamical assumption of the simulations, but we still consider this case useful, as it approximately sets a lower limit on the IC with respect to the synchrotron emission.

For each streamline we follow the same procedure: we identify along the line where the internal energy increases and the fluid velocity decreases. In these cells we assume there is a shock and therefore non-thermal particles are being injected, with a total energy corresponding to a fraction $\eta_{NT}=0.1$ of the internal energy increment. We note that there is no feedback of the non-thermal population on the hydrodynamics, i.e. we adopt a test particle approximation, and therefore the luminosity is linearly scalable with $\eta_{\rm NT}$ (incidentally, $\eta_{\rm NT}=1$ is consistent with $\hat\gamma=4/3$, but in this case the test-particle approximation fails for a radiatively efficient emitter). The particles are assumed to be injected with an energy distribution following a power-law of index $-2$, typical for shock acceleration, and with two cutoffs, one at high energies, derived assuming particle acceleration in a non-relativistic shock under Bohm diffusion \citep[e.g.][]{dru83,bp16}, and one at low energies. A significantly steeper power-law can be considered as another form of inefficiency accelerating high-energy emitting particles, equivalent to a small $\eta_{\rm NT}$-value. 
The low-energy cutoff is fixed to 1~MeV for simplicity, although our results are not very sensitive to this choice unless the low-energy cutoff is close to the energies of electrons emitting X-rays (mainly synchrotron) and gamma rays (mainly IC). Otherwise, only the normalization of the radiation spectrum will be slightly different.

\begin{figure}
    \centering
    \includegraphics[width=0.4\textwidth, angle=0]{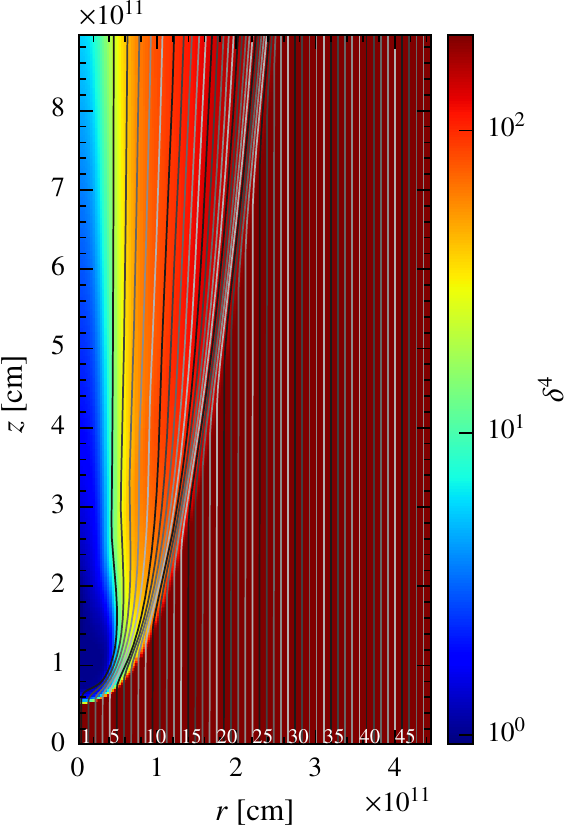}
    \caption[]{Doppler boosting factor for $\phi_{\rm obs}= 0^\circ$. The grey lines represent the computed streamlines, which are numbered.}
    \label{fig:dop}
  \end{figure}

\begin{figure}
    \centering
    \includegraphics[width=0.5\textwidth, angle=0]{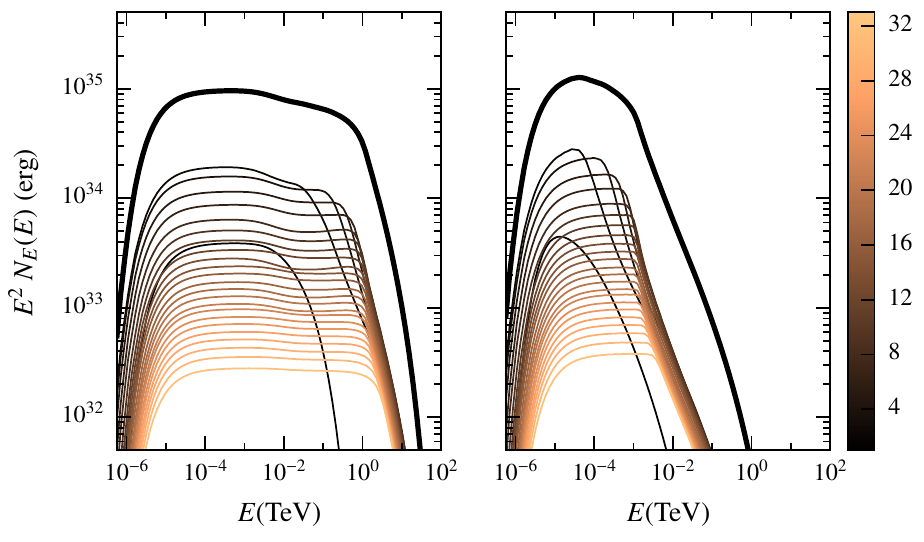}
    \caption[]{Compression-phase electron energy distribution for the individual streamlines (thin lines), and the total electron distribution (thick lines). The left panel corresponds to the low magnetic field case ($\chi_B=10^{-3}$), and the right panel to the high magnetic field case ($\chi_B=1$).}
    \label{fig:dist}
  \end{figure}

Once the injection of non-thermal particles is known, we let them evolve and propagate through the streamlines until they leave the simulation grid, point at which a steady-state is reached. The obtained energy distributions of the non-thermal particles for two $B$ cases, low and high, are shown in Fig. \ref{fig:dist}. Once the non-thermal population is characterized, one can compute the IC and synchrotron radiation in the FF, and later transform the spectral energy distribution (SED) of this radiation to the observer frame. For that, we multiply the photon energy by $\delta=1/\Gamma[1-\nu\cos(\phi_{\rm obs})]$ and the SED by $\delta^4$, where $\phi_{\rm obs}$ is the angle between the flow motion and the line of sight. The distribution of $\delta^4$ in the grid, adopting the most favourable case for detection ($\phi_{\rm obs}=0$), is shown in Fig.~\ref{fig:dop}. 

Note that in the low magnetic field case, the highest energy particles, despite still being confined to the interaction structure, could actually cross several streamlines before cooling significantly. This may affect the highest energy part of the spectrum to some extent, but for simplicity we assume that all electrons are attached to their respective streamlines as the bulk of the gamma-ray emission is radiated consistently with this assumption.

\subsection{Radiation results}

For simplicity, we consider in what follows the radiation results for a typical orbital phase: assuming a circular orbit, we considered an orbital phase  right in the middle between inferior and superior conjunction, i.e., the compact object is in the plane of the sky. This provides a sort of typical high-energy SED. A more detailed analysis of the overall spectra in the different explored cases, and for different orbital phases, is out of the scope of the paper, as we are interested here in the average behavior when the jet is present. We are also mostly interested in the 0.1--100~GeV band luminosity because this band is not strongly sensitive to parameters such as the maximum particle energy, and reacts smoothly to magnetic field and system-observer orientation changes, whereas the TeV band is very sensitive to all these factors through IC effects and gamma-ray absorption \citep[e.g.][gamma-ray absorption is included in the present work]{BosKha09}. The GeV luminosity is also available for Cyg~X-1 and Cyg~X-3, as these sources have been detected in this energy band. For low-to-moderate $B$-values, this band is also a good proxy for the source energetics and, unlike X-rays, is certainly of non-thermal nature.

\begin{figure}
    \centering
    \includegraphics[width=0.51\textwidth, angle=0]{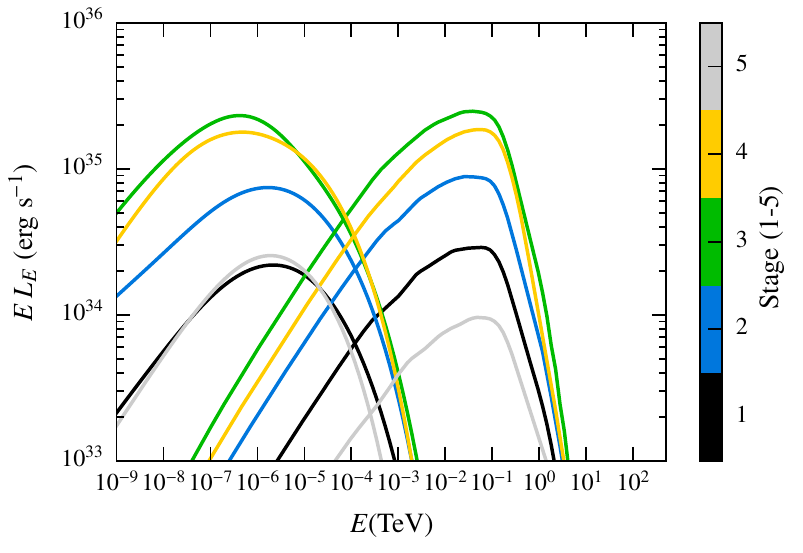}
    \caption[]{Spectral Energy Distributions for the 5 stages shown in Fig.~\ref{fig:dens}. Stage 1 corresponds to the compression phase, whereas stage 3 is the disruption phase; $\chi_B=10^{-3}$ is adopted, and $\phi_{\rm obs}=30^\circ$}
    \label{fig:seq}
  \end{figure}

Two illustrative stages of the clump evolution are considered when presenting the results of our computations: (i) the first stage, when the clump is compressed by the jet ram pressure (hereafter called compression phase), the longest and more stable phase of the clump evolution; and (ii) a later, shorter stage when the clump is disrupted (hereafter called disruption phase), presenting a larger shock and therefore higher non-thermal emission; both stages are shown in Fig.~\ref{fig:dens}, first and third panels starting from the left. We also present in Fig.~\ref{fig:seq} the SEDs of all 5 snapshots of the hydrodynamical flow shown in Fig.~\ref{fig:dens}. This illustrates how the high-energy emission varies with the clump evolution. The disruption phase is chosen as the maximum luminosity snapshot among those computed (i.e. number 3). From the point of view of the hydrodynamics, snapshot number 4 is characterized by a largest cross-section of the shocked jet, but the apparent non-thermal luminosity is higher for the hydrodynamic configuration shown in snapshot 3. This is likely related to a reduced Doppler boosting caused by streamline stronger deflection in 4; also, some additional weakening of the emission can be caused by the limited grid size.

\begin{figure}
    \centering
    \includegraphics[width=0.48\textwidth, angle=0]{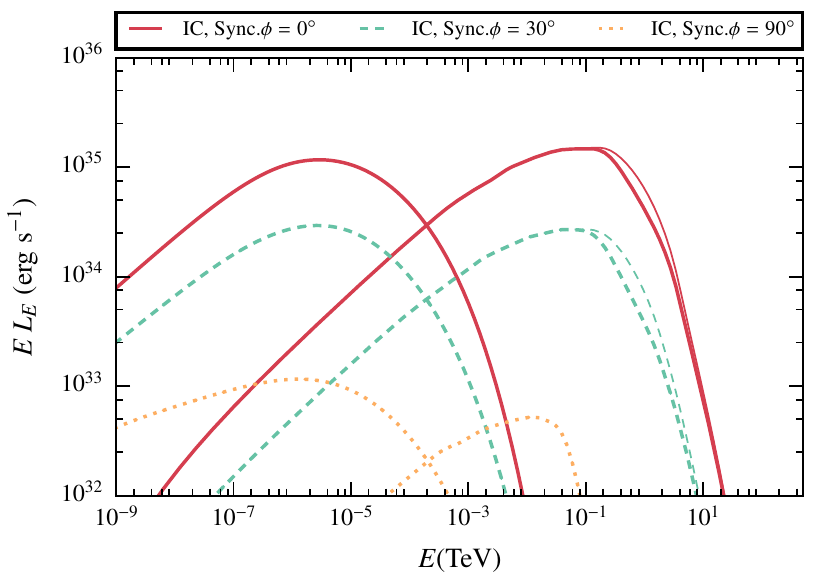}
    \includegraphics[width=0.48\textwidth, angle=0]{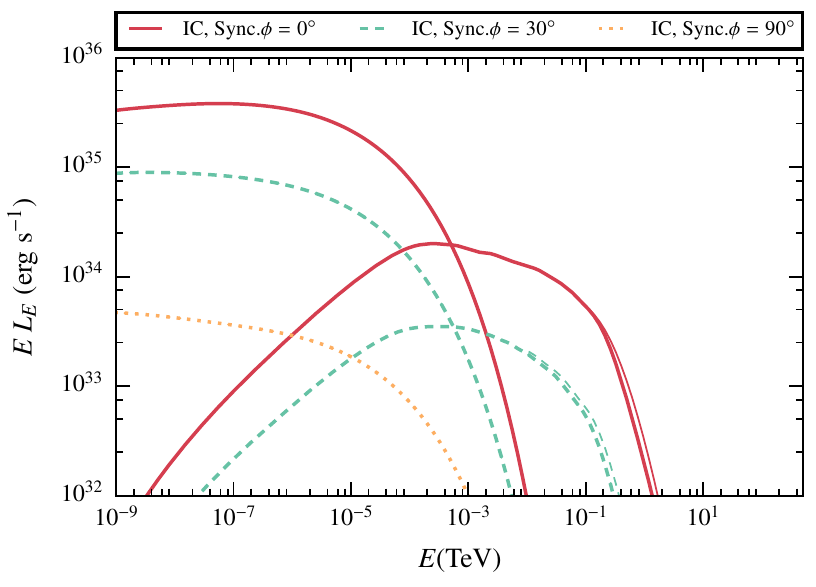}
    \includegraphics[width=0.48\textwidth, angle=0]{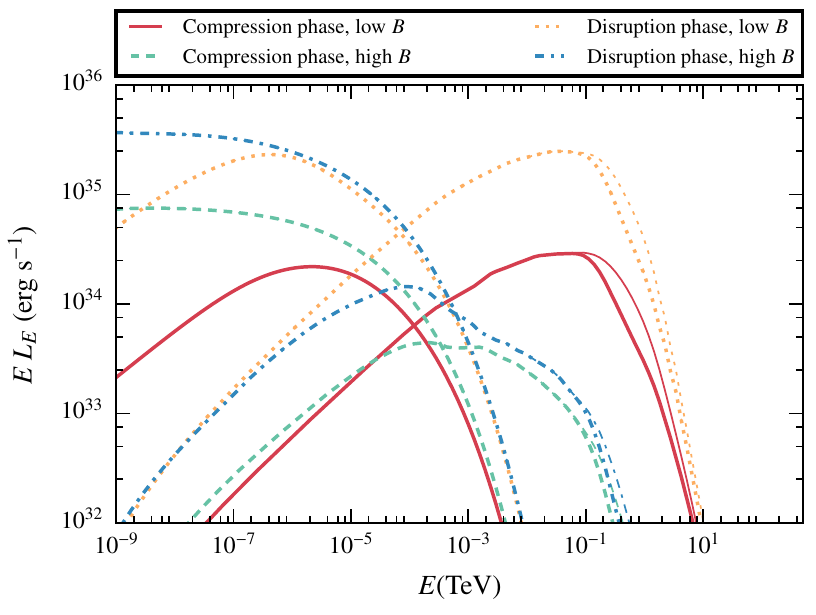}
    \caption[]{Top panel: Synchrotron and IC SEDs for the compression phase for $\phi_{\rm obs}=0$, 30, and $90^\circ$, and $\chi_B=10^{-3}$. The thin lines represent the emission without taking into account gamma-ray absorption due to electron-positron pair creation, i.e. the production SED. Middle panel: the same as for the top panel but for $\chi_B=1$. Bottom panel: a comparison between the SEDs of the compression and disruption phases.}
    \label{fig:com}
  \end{figure}

In the three panels of Fig.~\ref{fig:com}, we first focus on the compression phase and vary $B$, and then we compare the compression and disruption phases. It can be seen that, in the case a weak magnetic field, the transition from the compression to the disruption phase is accompanied by a flux increase by a factor of $\sim 5$.  In the high $B$ case, the emission enhancement is modest, within a factor of two. 

To illustrate the importance of the radiation losses, we can compare the energy injected per time unit in the form of non-thermal particles with that leaving the computational grid, after suffering energy losses, in the laboratory frame. For the compression phase, the total injected luminosity is $\approx 1.6\times 10^{35}\,(\eta_{\rm NT}/0.1)$~erg~s$^{-1}$ and about 33\% of the energy is kept by the particles when leaving the grid in the low $B$ case, being this percentage smaller ($\approx 15\%$) for the high $B$ case.
In that stage, particles lose a $\approx 49\%$ of the injected energy through adiabatic losses, which means $\approx 7.7 \times 10^{34}$ erg s$^{-1}$; the synchrotron+IC losses are $\approx 19$\% of the injected energy in the low $B$ case, which means $\approx 3 \times 10^{34}$~erg~s$^{-1}$ ($\approx 36\%$ and $\approx 5.8 \times 10^{34}$ erg s$^{-1}$ in the high $B$ case). In the disruption phase, the total injected luminosity is $\approx 10^{36}\,(\eta_{\rm NT}/0.1)$~erg~s$^{-1}$. In that stage, particles actually gain energy through adiabatic heating as in some streamlines the flow is compressed, approximately $\approx 2\times 10^{35}$ erg s$^{-1}$ ($\approx 20\%$ of the injected energy); the synchrotron+IC losses are $\approx 50$\% of the injected energy in the low $B$ case, which means $\approx 5\times 10^{35}$~erg~s$^{-1}$ ($\approx 95\%$ and $\approx 9.5\times 10^{35}$~erg~s$^{-1}$ in the high $B$ case). For the sake of discussion (see Sect.~\ref{sec:disc}), we provide in Table~\ref{tab:bands} the integrated luminosity in the range 0.1--100~GeV for the different cases studied here: the compression and disruption phases, for the low and the high $B$ cases, and $\phi_{\rm obs}=30^\circ$, given that this observing angle may be representative of both Cyg~X-1 and Cyg~X-3.

\begin{table}
\caption{Values of the integrated emission in the 0.1--100~GeV band, given in erg~s$^{-1}$, considering an observer angle of $\phi_{\rm obs}=30^\circ$. The compression and disruption states are computed based on the first and third snapshots in Fig.~\ref{fig:dens}, respectively.}
\label{tab:bands}
\centering
\begin{tabular}{ccc}
\hline\hline
    Stage       &     Low $B$ ($\chi_B=10^{-3}$)       &        High $B$ ($\chi_B=1$)   \\
        \hline
      Compression phase   & $ 1.30  \times 10^{35}  $             & $ 2.15  \times 10^{34}  $\\
\hline
      Disruption phase    & $ 1.12  \times 10^{36}  $            & $ 4.71  \times 10^{34}  $\\
\hline
\end{tabular}

\end{table}

\section{Collective effects of a clumpy wind} \label{sec:results}

Clumping is universal in massive star winds: these winds are stochastically inhomogeneous, believed to be composed by a hierarchy of clumps, with few large ones and increasingly many more small ones, being the clump size and mass distributed as a power-law \citep{MofCor09}.  The X-ray spectrum of single and binary massive stars is compatible with this picture of dominant small-scale clumps and rarer large clumps \cite{Mof08}. 

For simplicity, we consider the clumps to be spherical, and neglect vorosity
\citep[i.e. porosity in velocity space, see e.g.][]{mv11}, which in the present context can be considered as a minor effect.
An empirical number density distribution of clumps with radius $R_\mathrm{c}$ is adopted: 
\begin{equation}
    n(R_\mathrm{c}) = \frac{\mathrm{d}N}{\mathrm{d}R_\mathrm{c} \mathrm{d}V} = n_0 {R_\mathrm{c}}^{-\alpha},
\end{equation}
with clump radii ranging $R_\mathrm{c, min} < R_\mathrm{c} < R_\mathrm{c, max}$.
The value of $R_\mathrm{c, min}$ can be considered close to the Sobolev length,
$R_{\mathrm{Sob}}\approx 0.01 R_*$ \citep[e.g.][]{OwoCoh06}, while
the clump size is at most of the order of $R_*$ \citep{lhf10}, although it may be significantly smaller. As clumps propagate from the region where they form, at $\sim 1-2\,R_*$ from the star centre, they can grow linearly with the stellar distance or somewhat slower for a slab geometry, but they can be also broken down by instabilities \citep[see the discussion at the end of sect.~3.3 in][]{Bos13}.

Assuming that all clumps have the same density and that the 
inter-clump medium is void, the clump volume filling factor is simply 
$f = \chi^{-1}$ \citep[e.g.][]{hof08}, and the distribution 
function should meet the following normalization condition:
\begin{equation} \label{eq:clump_norm}
    \frac{4\pi}{3} \int_{R_\mathrm{c,min}}^{R_\mathrm{c,max}} R^3_\mathrm{c} \, n(R_\mathrm{c}) \, \mathrm{d}R_\mathrm{c} = f\,.
\end{equation}
Fixing $R_\mathrm{c,min}=0.01 R_*$ and $R_\mathrm{c,max}=0.5 R_*$, 
we can solve Eq.~(\ref{eq:clump_norm}) to obtain the normalization constant $n_0$. 

Whether clump-jet interactions appear as a transient or as a persistent phenomenon is determined
by the duty-cycle ($DC$) of these events. To determine $DC$ we need to estimate the jet penetration rate of the clumps, $\dot{N}$, and their lifetime,
$t_\mathrm{c}$ \citep{Bos13}. The jet crossing time will be more relevant than $t_\mathrm{c}$ for jet powers $R_{\rm c}/R_{\rm j}$ times the limit provided in Eq.~(\ref{eq:Ljet}). In this work we are interested in bright sources, and therefore the jet power will be assumed to 
be close to the limit provided in Eq.~(\ref{eq:Ljet}). In fact, this is likely the case in the powerful jet sources Cyg~X-3 and Cyg~X-1.

The jet subtends a certain solid angle, $\Omega$, as seen from the optical star. To interact with the jet, a clump must propagate within this solid angle. Considering a conical jet with $R_\mathrm{j}=\theta_\mathrm{j}z_{\rm j}$
and that the clump enters the jet roughly between 
$0.5 R_\mathrm{orb}\lesssim z\lesssim 1.5 R_\mathrm{orb}$ (so $\Delta z\sim R_{\rm orb}$), we obtain $\Omega\approx \theta_j/2=0.05$.
The differential rate of arrival to the jet for clumps of radius $R_\mathrm{c}$ can be estimated 
as $\mathrm{d}\dot{N} = \Omega d^2 v_{\rm w} n(R_{\rm c})\mathrm{d}R_{\rm c}$, where for simplicity $d\approx\sqrt{R_{\rm orb}^2+z_{\rm j}^2}$.
The differential $DC$ for those clumps is $\mathrm{d}DC = t_\mathrm{c} {\rm d}\dot{N}$, with $t_\mathrm{c} \sim R_\mathrm{c}/v_\mathrm{sh}$, which we approximate here to $t_{\rm c}=R_{\rm c}/v_{\rm w}$ as we deal with powerful jets\footnote{This assumption makes the estimate of the luminosity for collective interactions more conservative.}. From all this, the following estimate for the arrival of clumps with radius between $R_1$ and $R_2$ can be obtained:
\begin{equation} \label{eq:DC}
    DC(R_1,R_2)= \Omega d^2 n_0 \int_{R_1}^{R_2} R_\mathrm{c}^{-\alpha+1}\mathrm{d}R_\mathrm{c} \approx \frac{\theta_{\rm j}}{2} d^2 n_0 \int_{R_1}^{R_2} R_\mathrm{c}^{-\alpha+1} \mathrm{d}R_\mathrm{c}. 
\end{equation}

We explore different values of the power-law index: $2.5 \leqslant \alpha \leqslant 6$.  
Values of $\alpha \ge 4$ imply that small clumps dominate the wind mass. Interestingly, as the energy emitted by one clump-jet interaction is 
expected to be $\propto t_{\rm c}\times R_{\rm c}^2 \propto R_{\rm c}^3$, one also obtains that for $\alpha > 3$ the non-thermal radiation
will be dominated by the smallest clumps that can enter into the jet, i.e. those with $R_{\rm c}\sim R_{\rm 0}$. On the other hand, values of 
$\alpha\approx 2.5$ are in accordance to the values inferred for WR stars \citep{Mof08}\footnote{The value given is
actually $N(m) \propto m^{\gamma}$, with $\gamma = 1.5 \pm 0.1$. Assuming constant
density, spherical clumps, one derives $\alpha\approx 2.5$.}. 

The non-thermal luminosity of the clump-jet collective interactions depends on the dominant clump size; gamma rays can be mostly produced by small or large clumps. The $DC$ may be dominated by small clumps, and still the luminosity may be dominated by the largest ones. Nevertheless, the simulations show that clumps significantly larger than $R_0$ cannot increase considerably the luminosity output, and may even radiate less than smaller clumps interacting with the jet. The reason is that, unless $(\theta_\mathrm{j}/0.1)(10/\chi)$ is well below one (see Eq.~\ref{eq:clump_min}), larger clumps will likely disrupt the whole jet, strongly reduce the effect of Doppler boosting, and might even switch off particle acceleration. For the parameters adopted in this work, it seems therefore natural to consider clumps of radii between $\sim R_0$ and few times larger.
 
  \begin{figure}
    \centering
    \includegraphics[width=0.4\textwidth, angle=270]{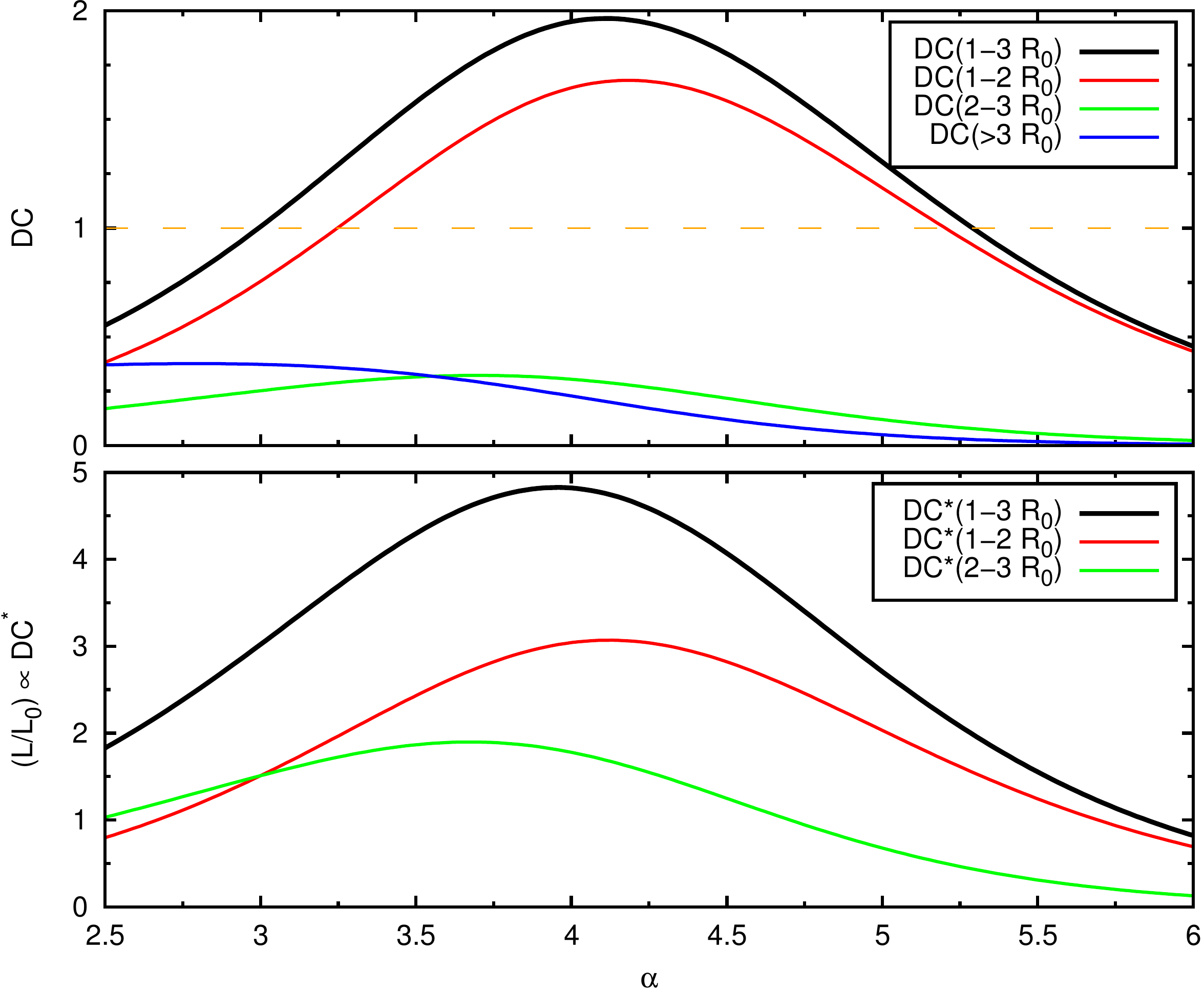}
    \caption[]{Top panel: Duty-cycle in the radius ranges $1-2 \,R_0$, $2-3 \,R_0$, $1-3 \,R_0$, and $>3 \,R_0$ for a wind clump number distribution $n \propto {R_\mathrm{c}}^{-\alpha}$. A value $\alpha \approx 5$ represents a wind dominated by small clumps, whereas $\alpha_2 \approx 2.5$ is an observational value for WR stars; $R_0$ is the minimum clump size given by Eq.~(\ref{eq:clump_min}). Bottom panel: Duty-cycle weighted by the effective section of the shock, and therefore its potential luminosity, shown in units of the luminosity $L_0$ produced in the interaction of one clump of $R=R_0$ with the jet. The case $>3\,R_0$ is not shown as it may imply jet disruption.}
    \label{fig:DCs}
  \end{figure}  
  
The top panel of Fig.~\ref{fig:DCs} shows the $DC$ dependence on $\alpha$, whereas the bottom panel shows the same but for the average luminosity of the collective clump-jet interactions. We have focused on $1\,R_0<R_{\rm c}<3\,R_0$, splitting this range into $1\,R_0<R_{\rm c}<2\,R_0$ and $2\,R_0<R_{\rm c}<3\,R_0$, although a curve for $R_{\rm c}>3\,R_0$ is also shown. The radius $3\,R_0$ has been considered as the upper-limit for the clumps to be relevant from the radiation point of view in the context of this work, that is $(\theta_\mathrm{j}/0.1)(10/\chi)\sim 1$. The results obtained allow the derivation of a rather robust conclusion: clumps with $R_{\rm c}>R_0$ will be always present inside the jet unless the $\alpha$ parameters deviate strongly from the expected values. Also, the averaged total luminosity should be a factor of a few larger than the estimate obtained for a jet interacting with one clump of radius $R_0$ (again unless $\alpha$ is in the extremes of the explored range). We studied cases with different $R_\mathrm{c,max}$-values (not shown here), and found that our conclusions hold for a wide range of $R_\mathrm{c,max}\sim 0.1-1 R_*$.

For jet powers well below the value given in Eq.~(\ref{eq:Ljet}), the clump-jet interaction luminosity is $\propto L_{\rm j}$ and therefore lower than the reference case, but the event duration is longer by $\propto v_{\rm sh}^{-1}\propto L_{\rm j}^{-1/2}$ (recall that $v_{\rm sh}$ must be used instead of $v_{\rm w}$, see above). Thus, to first order, the decrease in the radiation luminosity will effectively be $\propto L_{\rm j}^{1/2}$, although a more accurate relation should be derived numerically. 

If clumps with $R_{\rm c}>3\,R_0$ are present and $DC\gtrsim 0.5$, the jet will be likely disrupted most of the time in the region of interest, $z_{\rm j}\sim R_{\rm orb}$. As noted, this is expected to significantly reduce the effects of Doppler boosting, and potentially might even switch off particle acceleration. In the scenario explored, i.e. $(\theta_\mathrm{j}/0.1)(10/\chi)\sim 1$, a value $DC\sim 0.5$ for $R_{\rm c}>3\,R_0$ will be achieved only for $\alpha\lesssim 3$ and $R_{\rm c,max}\sim 0.2\,R_*$. This sets limits on the wind properties that allow clump-jet interactions to produce significant gamma-ray emission. These are however somewhat extreme, possibly unrealistic, wind parameters, in particular concerning the constraint on $\alpha$, as in that case the wind mass would be dominated by the largest clumps.

Finally, let us explore situations different from $(\theta_\mathrm{j}/0.1)(10/\chi)\sim 1$: 
\begin{enumerate}[label=(\roman*)]
 \item {Fixing $(\theta_\mathrm{j}/0.1)\sim 1$: Sources with $(10/\chi)>1$ would imply that $R_0$ would be larger, and there may be no clumps big enough to cross the shocked wind surrounding the jet. In addition, a $\chi$ well below 10 would mean a rather homogeneous wind. On the other hand, for $(10/\chi)<1$, $R_0$ would be smaller; all the clumps may be able to reach the jet, and they may be also dense enough to fully enter inside. This would have the disrupting effect of a wind with most of its mass in large, penetrating clumps \citep[this kind of scenario is simulated in][]{PerBos12}. Additionally, the number of clumps would be lower by $\chi^{-1}\propto f$ (see Eq.~(\ref{eq:clump_norm})), but because $t_\mathrm{c}$ is actually $\propto f^{-1/2} \propto \chi^{1/2}$, the radiation luminosity would be $\propto f^{1/2} \propto \chi^{-1/2}$ lower.}
 \item {Fixing $(10/\chi)\sim 1$: Taking $(\theta_\mathrm{j}/0.1)<1$ would imply that smaller clumps could reach the jet but the latter would be denser, making clump penetration more difficult. Also, jet penetration would be less frequent ($DC \propto \Omega \propto \theta_j$). On the other hand, if $(\theta_\mathrm{j}/0.1)>1$, jet penetration would be significantly easier and more frequent, but clumps should be larger to first cross the shocked wind surrounding the jet.}
\end{enumerate}

\section{Discussion and summary} \label{sec:disc}

\begin{figure}
  \centering
  \includegraphics[width=0.5\textwidth]{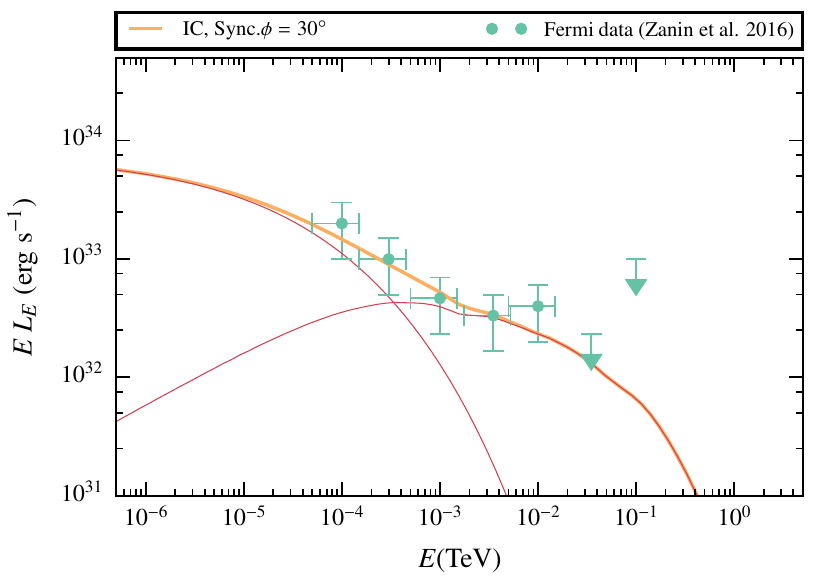}
  \caption[] {Synchrotron and IC SEDs (thin lines, red) computed for the  compression phase and the sum of the two contributions (thick, orange line) plotted together with the Fermi data of Cyg~X-1 published in \cite{zf16}. In this case, we have adopted a strong magnetic field ($\chi_{\mathrm B} = 0.5$) and fixed the acceleration efficiency to $\eta_{\rm NT} = 0.01$. The observing angle is $\phi_\mathrm{obs} = 30^\circ$.}
  \label{fig:cyg-fermi}
\end{figure}

For typical values of the clumping factor of massive star winds, and of the jet geometry and power, we obtain that clumps of intermediate size, say a few \% of the stellar radius, can overcome the shocked wind surrounding the jet, and penetrate into the latter. For $\chi\sim 10$, clumps can already sustain the jet impact long enough for fully penetrate into the jet, allowing for an dynamically strong interaction. Under such circumstances, and assuming moderate acceleration efficiencies, say $\eta_{\rm NT}\sim 0.01-0.1$, we predict significant gamma-ray luminosities for galactic sources at few kpc distances. For mildly relativistic jets, the impact of Doppler boosting is non-negligible even for relatively large jet viewing angles, $\sim 30^\circ$. If $\Gamma_{\rm j}$ were higher, clump-jet interactions would be detected for viewing angles within a relatively narrow cone around the jet orientation, whereas $\Gamma_{\rm j}\rightarrow 1$ would mean lower luminosities, although potentially still detectable if non-thermal efficiencies were high.

In what follows we focus on the possibility of explaining the persistent GeV emission detected from Cyg~X-1 and Cyg~X-3 during the low-hard state and GeV activity periods, respectively. The calculations presented in Sect.~\ref{sec:simulation} for one clump-jet interaction were carried out for a system with similar properties to those of Cyg~X-1. For Cyg~X-3, the results would be similar, but $R_{\rm orb}$ and thus $R_0$ would be $\sim 10$ times smaller, and the jet power and therefore the non-thermal emission $\sim 10$ times higher \citep[see][and references therein for a comparison of these two sources]{yzh16}. The IC luminosity in the high $B$ case would be lower though with respect to the injected luminosity because of a more compact emitting region. From the one-clump interaction properties, one can extrapolate the characteristic gamma-ray luminosity in the context of collective clump-jet interactions. 

The collective clump-jet interaction luminosity, in the GeV ($0.1-100$~GeV) range, calculated averaging over one orbit\footnote{The GeV luminosity was also computed in inferior and superior conjunction so that the estimate was representative.}, and the computed clump evolution, for a jet inclination with the line of sight of $\phi_\mathrm{obs}=30^\circ$, is $\sim 10^{35}\,(\eta_{\rm NT}/0.1)$~erg~s$^{-1}$ for Cyg~X-1 ($\sim 10^{36}\,(\eta_{\rm NT}/0.1)$~erg~s$^{-1}$ for Cyg~X-3), with an uncertainty of a factor $\sim 0.5$--2 (including the high and the low $B$ cases in that range). Assuming that the jet in Cyg~X-1 is perpendicular to the orbital plane, its inclination with the line of sight is $\phi_\mathrm{obs}\sim 30^\circ$ \citep{oma11}. In Cyg~X-3, the jet inclination may be similar or even smaller (see, e.g., \citealt{mrh01,dch10}; see however \citealt{mpp01}). Section~\ref{sec:results} shows that $DC\sim 1$ are expected. Therefore, taking into account that the GeV luminosity in Cyg~X-1 and Cyg~X-3 are $\approx 5\times 10^{33}$ \citep{zf16} and $\approx 3\times 10^{36}$~erg~s$^{-1}$ \citep{faa09}, respectively, one can derive the required non-thermal efficiency to be $(\eta_{\rm NT}/0.1)\sim 0.05$ for Cyg~X-1, and $\sim 3$ for Cyg~X-3 (for $\phi_\mathrm{obs}=30^\circ$; $\sim 0.5$ for $\phi_\mathrm{obs}\sim 0^\circ$). 

For Cyg~X-1 the observational constraints only require a very modest non-thermal fraction ($\sim 1$\%). A comparison with Fermi data published in \cite{zf16} is shown in Fig.~\ref{fig:cyg-fermi}, where a rather strong magnetic field ($\chi_{\mathrm B} = 0.5$) and a fixed acceleration efficiency ($\eta_{\rm NT}=0.01$) were needed to reproduce the observational data, assuming $DC=1$. We note that this is a simplified model, but it is interesting that even so it can approximately match the Fermi data. A similar toy-application of our model to Cyg~X-3 could be also performed (not shown here), for similar parameters to those of Cyg~X-1 but a much larger non-thermal fraction. In this case, the energetics is however rather demanding ($\sim 30$\%). Nevertheless, the presented calculations are obtained fixing $\Gamma_{\rm j}=2$, while in fact different $\Gamma_{\rm j}$-values are possible, and despite the relation between $\Gamma_{\rm j}$ and the Lorentz factor of the shocked jet flow is non-trivial (it depends on the postshock flow re-acceleration), slightly faster jets could alleviate the tight energetics for the Cyg~X-3-like scenario. In addition, adopting a low magnetic field and the (probable) possibility of $DC$ of a few would relax further the energetic constraints. Finally, the numerical calculations carried out are also likely to underestimate the non-thermal emission because the computational grid encloses a relatively small region, and part of the radiation, and the last stages of the clump evolution, are not accounted for (see Sect.~\ref{sec:simulation}). 

Values of $DC\ll 1$ would lead to flares rather than to a smoother clump-jet interaction continuum. This requires a much higher degree of inhomogeneity than the assumed in most of this work, as strong changes in $R_\mathrm{c,min}$ and $R_\mathrm{c,max}$ are not possible, and $DC$ is not very sensitive to $\alpha$. However, as noted in Sect.~\ref{sec:results}, the radiation luminosity is $\propto \chi^{-1/2}$, meaning that for the same parameters, very inhomogeneous winds interacting with jets will be on average less radiatively efficient than moderately inhomogeneous ones.

\begin{acknowledgements}
We acknowledge support by the Spanish Ministerio de Econom\'ia y Competitividad (MINECO) under grants
AYA2013-47447-C3-1-P, AYA2016-76012-C3-1-P, and MDM-2014-0369 of ICCUB (Unidad de Excelencia 'Mar\'ia de Maeztu'), and the Catalan DEC grant 2014 SGR 86. 
This research has been supported by the Marie Curie Career Integration Grant 321520. V.B-R. also acknowledges financial support from MINECO and European Social Funds through a Ram\'on y Cajal fellowship.
This work is supported by ANPCyT (PICT 2012-00878). X.P-F. also acknowledges financial support from Universitat de Barcelona and Generalitat de Catalunya under grants APIF and FI (2015FI\_B1 00153), respectively. G.E.R. and S.dP. are supported by grant PIP 0338, CONICET. D.K. acknowledges support by the Russian Science Foundation under grant 16-12-10443. 
\end{acknowledgements}

\bibliographystyle{aa} 
\bibliography{ALLreferences} 

\end{document}